\documentclass[11pt]{article}
\usepackage{amssymb}
\usepackage{epsfig}
\usepackage{verbatim}
\begin{document}
\title{ Full one-loop supersymmetric corrections to charged Higgs boson pair production in $\gamma\gamma$  collisions
 \footnote{Supported by National Natural Science Foundation of China.}} \vspace{3mm}

\author{{ Wang Lei$^{2}$, Jiang Yi$^{2}$, Ma Wen-Gan$^{1,2}$, Han Liang$^{2}$, and  Zhang Ren-You$^{2}$} \\
{\small $^{1}$ CCAST (World Laboratory), P.O.Box 8730, Beijing 100080, P.R.China} \\
{\small $^{2}$ Department of Modern Physics, University of Science and Technology} \\
{\small of China (USTC), Hefei, Anhui 230027, P.R.China}}
\date{}
\maketitle \vskip 20mm
\begin{abstract}
The complete one-loop electroweak corrections to charged Higgs
boson pair production in $\gamma\gamma$ collision mode at linear
colliders in the minimal supersymmetric standard model (MSSM), are
calculated in this paper. We discuss the dependence of the
corrections to the subprocess $\gamma\gamma \to H^{+}H^{-}$ on the
related parameters, such as the colliding energy, charged Higgs
boson mass $M_{H^{\pm}}$ and some supersymmetric parameters $\tan
\beta$, $M_{SUSY}$ and gaugino mass parameter $M_2$. We find that
the corrections generally reduce the Born cross sections and the
relative one-loop corrections to both the subprocess typically in
the range of $-10\%$ to $-30\%$. We also present the numerical
results at the SPS1a' point from the SPA project. We conclude that
the full one-loop electroweak corrections to subprocess $\gamma
\gamma \to H^{+}H^{-}$ and the parent process $e^+e^- \to \gamma
\gamma \to H^{+}H^{-}$ are significant and therefore should be
considered in precise analysis of charged Higgs boson pair
productions via $\gamma\gamma$ collision at future linear
colliders.

\end{abstract}
\vskip 20mm {\large\bf PACS: 12.15.LK, 12.60.Jv, 12.38.Bx,
14.80.Cp}

\vfill \eject \baselineskip=0.36in
\renewcommand{\theequation}{\arabic{section}.\arabic{equation}}
\renewcommand{\thesection}{\Roman{section}}
\newcommand{\nb}{\nonumber}
\makeatletter      
\@addtoreset{equation}{section}
\makeatother       
\section{Introduction}
\par
New physics beyond the standard model (SM) has been intensively
studied over the past years\cite{haber}. Most extensions of the SM
require an electroweak symmetry breaking sector, which is composed
of two scalar isospin-doublets, and the charged Higgs bosons are
part of its physical spectrum at the weak scale. The minimal
supersymmetric standard model (MSSM) is one of the typical
example. From the phenomenological point of view, if a light
neutral Higgs boson was found, it is still very hard to tell which
model it belongs to, since the fundamental properties of such a
particle (quantum number, couplings, branching ratios, etc.) are
almost the same in some models, e.g. in the SM and in the
'decoupling regime' of the MSSM (i.e., when $M_{A^0}>200~GeV$,
$M_{A^0}\sim M_{H^0}\sim M_{H^{\pm}}>>M_{h^0}$). While the
discovery of the charged Higgs boson is an unambiguous signature
of existing new physics beyond the SM.

\par
Historically, a lot of effort has been invested in the charged
Higgs boson pair production at the future colliders, such as the
CERN Large Hadron Collider (LHC), Tevatron, and the proposed
linear colliders (LC): NLC\cite{NLC}, JLC\cite{JLC},
TESLA\cite{TESLA} and CLIC\cite{CLIC}.
Refs.\cite{jiang}\cite{bbhh} presented the calculations of the
charged Higgs boson pair productions at hadron colliders in
different important production channels. It shows that the
production cross section can reach few femto-bar. Linear colliders
can also produce the charged Higgs pair with larger production
rate, because the process can occur at the tree level and is not
suppressed by the light Yukawa couplings. Furthermore, the
signature of event at LC is much cleaner than that produced at
hadron colliders. With the help of high integrated luminosity, the
precise measurement at LC for probing the new physics is possible.
Therefore, the theoretical calculations beyond the tree-level are
necessary in studying the charged Higgs boson productions. In
Ref.\cite{arhrib}, the process $e^+e^-\to H^+H^-$ involving
one-loop fermion and sfermion corrections has been studied, it
points out that the corrections are about $-10\%$ in a wide range
of parameter space of the MSSM. Ref.\cite{jaume} gives the
complete one-loop electroweak corrections to the cross section of
the process $e^+e^-\to H^+H^-$ in the THDM as well as the MSSM. It
shows that the corrections vary in the range between $-15\%$ and
$10\%$. The O($\alpha m_t^2/m_W^2$) Yukawa corrections to the
process $e^+e^- \to \gamma\gamma\to H^+H^-$ in the THDM were
studied in Ref.\cite{ma3}. Ref.\cite{zhu} presents the squarks
one-loop corrections to the process $e^+e^- \to \gamma\gamma\to
H^+H^-$ in the MSSM. It says that the relative corrections are
from $-25\%$ to $25\%$. From the previous works which deal with
the complete one-loop corrections to the new particle production
processes, we know that the detailed study of the one-loop
electroweak corrections for those processes at a very high
colliding energy is necessary. An electron-positron LC can be
designed to operate in either $e^+e^-$ or $\gamma\gamma$ collision
mode. $\gamma\gamma$ collision is achieved by using Compton
backscattered photons in the scattering of intense laser photons
on the initial polarized $e^+e^-$ beams\cite{Com}. Normally, the
cross section for $\gamma \gamma \to H^+ H^-$ is larger than that
of $e^+e^- \to H^+H^-$ due to the fact that the production rate in
$e^+e^-$ collision mode is s-channel suppressed. In this paper, we
present the calculations of the full one-loop radiative
corrections to the process $e^+e^- \to \gamma\gamma\to H^+H^-$ in
the MSSM. The paper is organized as follows. In Sec.II. we discuss
the LO results of the subprocess $\gamma\gamma\to H^+H^-$. In
Sec.III. we give the analytical calculations of the full one-loop
corrections. The numerical results and discussions are presented
in Sec.IV. Finally, we give a short summary.

\par
\section{The Leading Order Cross Section of subprocess $\gamma\gamma\to H^+H^-$}

\par
We denote the subprocess $\gamma\gamma\to H^+H^-$ as
\begin{equation}
\gamma(p_1)+\gamma(p_2) \to H^+(p_3)+H^-(p_4),
\end{equation}
where $p_{1},~p_{2}$ and $p_{3},~p_{4}$ represent the four-momenta
of the incoming partons and the outgoing particles, respectively.

\begin{figure}[htb]
\centering
\includegraphics{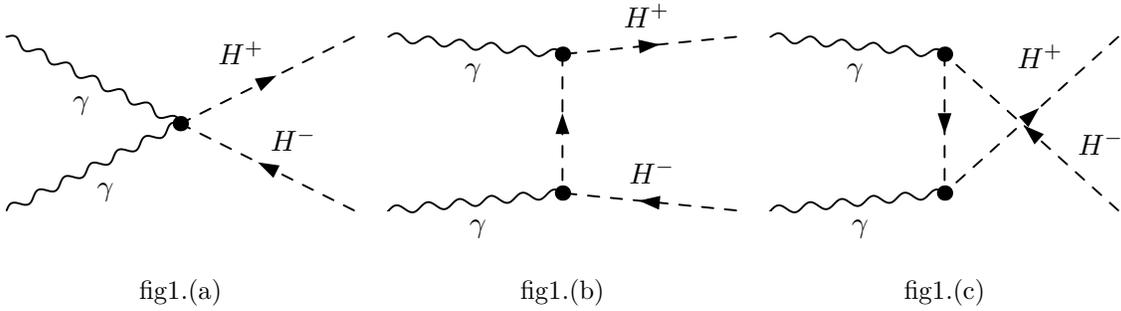}
\caption{The leading order diagrams for the $\gamma\gamma \to
H^+H^-$ subprocess.}
\end{figure}

\par
The Feynman diagrams for the subprocess $\gamma\gamma\to H^+H^-$
at the leading order(LO) are shown in Fig.1. There are three
Feynman diagrams for this subprocess at the tree-level. The
corresponding tree-level amplitude of the subprocess $\gamma
\gamma \to H^+H^-$ can be represented as
\begin{equation}
 M_0 = M_0^{\hat{t}} + M_0^{\hat{u}} + M_0^{\hat{q}}
\end{equation}
where $M_0^{\hat{t}}$, $M_0^{\hat{u}}$ and $M_0^{\hat{q}}$
represent the amplitudes arising from the t-channel, u-channel and
quartic coupling diagrams, respectively. The explicit expressions
can be written as
\begin{equation}
M_0^{\hat{t}} = \frac{i e^2}{\hat{t} - m_{H^\pm}^2 }
                (p_2-2 p_4)_{\nu}\epsilon_{\nu}(p_2)
                (p_2+p_3-p_4)_{\mu} \epsilon_{\mu}(p_1),
\end{equation}

\begin{equation}
M_0^{\hat{u}} = M_0^{\hat{t}} ~ (\hat{t} \to \hat{u},~~p_2 \to
p_1),~~~ M_0^{\hat{q}} = 2 i e^2 g^{\mu\nu} \epsilon_{\mu}(p_1)
\epsilon_{\nu}(p_2).
\end{equation}
The Mandelstam variables $\hat{t}$, $\hat{u}$ and $\hat{s}$ are
defined as $\hat{t}=(p_1 - p_3)^2,~\hat{u}=(p_1 - p_4)^2,~
\hat{s}=(p_1 + p_2)^2 $.

\par
Then the LO cross section for the subprocess $\gamma\gamma \to
H^+H^-$ is obtained by using the following formula:
\begin{eqnarray}
\label{folding} \hat{\sigma}^0(\hat{s}, \gamma\gamma \to H^+H^-) =
\frac{1}{16 \pi \hat{s}^2} \int_{\hat{t}_{min}}^{\hat{t}_{max}}
d\hat{t}~ \overline{\sum} |M^0|^2,
\end{eqnarray}
where $\hat{t}_{max,min}= (M^2_{H^\pm}- \frac{1}{2} \hat{s}) \pm
\frac{1}{2}\sqrt{\hat{s}^2- 4 M^2_{H^\pm} \hat{s}} $. The
summation is taken over the spins of initial and final states, and
the bar over the summation denotes averaging over the spins of
initial partons.

\par
\section{The Calculation of the Full One-loop Corrections to the subprocess $\gamma\gamma \to
H^+H^-$}

\par
In our calculations we use the t'Hooft-Feynman gauge. In the
calculation of one-loop diagrams we adopt the definitions of
one-loop integral functions in Ref.\cite{s13}. In order to control
the ultraviolet(UV) divergences, we take the dimensional reduction
($DR$) regularization scheme, which is commonly used in the
calculation of the electroweak correction in the framework of the
MSSM as it preserves supersymmetry at least at one-loop
order\cite{copper}. In doing renormalization we use
on-mass-shell(OMS) scheme\cite{ross}. The Feynman diagrams and
their amplitudes are automatically generated by using ${\it
FeynArts}~3$ package\cite{FA3}.

\par
\subsection{Virtual Electroweak One-loop Corrections }

\par
There are total 570 one-loop Feynman diagrams for the subprocess
$\gamma\gamma \to H^+H^-$ in the MSSM, and we can classify them
into four groups: self-energy, vertex, box diagrams and
counter-term diagrams. Let's consider the counter terms at first.
The Higgs potential in the MSSM can be divided into four parts
\begin{eqnarray}
\label{Higgs potential} V_H = V_H^{(1)} + V_H^{(2)} + V_H^{(3)} +
V_H^{(4)}
\end{eqnarray}
which represent the linear, quadratic, cube and quartic terms
respectively. The linear and quadratic terms can be expressed as
\begin{eqnarray}
\label{V_H^1 and V_H^2neu}
V_H^{(1)} &=& T_H H^0 + T_h h^0, \nb \\
V_H^{(2)\rm{charge}} &=& \frac{1}{2}
        \left(
        \begin{array}{cc}
        G^+ H^+
        \end{array}
        \right)^\dagger
    \left(
        \begin{array}{cc}
        b_{GG} & b_{GA} \\
    b_{GA} & M_{H^{\pm}}^2 + b_{AA}
        \end{array}
    \right)
    \left(
    \begin{array}{c}
    G^+ \\
    H^+
    \end{array}
    \right),
\end{eqnarray}
where
\begin{eqnarray}
\label{b function}
 b_{GG} &=& \frac{g}{2
m_W} [ T_H \cos(\alpha - \beta)
                           - T_h \sin(\alpha - \beta) ], \nb \\
b_{AA} &=& \frac{g}{m_W \sin2\beta} [ T_H (\sin^3\beta \cos\alpha
+ \cos^3\beta \sin\alpha)
                                    + T_h (\cos^3\beta \cos\alpha - \sin^3\beta \sin\alpha) ], \nb \\
b_{GA} &=& \frac{g}{2 m_W} [ T_H \sin(\alpha - \beta)
                           + T_h \cos(\alpha - \beta) ].
\end{eqnarray}

\par
We use the following definitions of the renormalization constants
related in our calculation as,
$$
 e_0=(1+\delta Z_e)e,~~~ A_0=\frac{1}{2} \delta
Z_{AZ}Z+(1+\frac{1}{2}\delta Z_{AA})A
$$
$$
T_{H,0}=T_{H}+\delta T_{H},~~~T_{h,0}=T_{h}+\delta T_{h}
$$
\begin{equation}
\label{counterterm}
 M_{H^{\pm},0}^2=M_{H^{\pm}}^2+\delta   M_{H^{\pm}}^2, ~ ~ ~
H^+_0=\frac{1}{2} \delta Z_{H^+G^+}G^++(1+\frac{1}{2}\delta
Z_{H^+H^+})H^+
\end{equation}
With the on-mass-shell conditions and tadpoles renormalization
condition $\hat{T}=T+\delta T=0$, we can obtain the renormalized
constants expressed as,

\begin{equation}
\label{ct1}
\delta Z_{AA}= -\tilde{Re}\frac{\partial
\Sigma_T^{AA}(p^2)}{\partial p^2}|_{p^2=0},~~~ \delta Z_{ZA}=
2\frac{\tilde{Re}\Sigma_T^{ZA}(0)}{m_Z^2},
\end{equation}

\begin{equation}
\label{ct2} \delta Z_e = -\frac{1}{2} \delta Z_{AA} +
\frac{s_W}{c_W}\frac{1}{2} \delta Z_{ZA}
           = \frac{1}{2} \tilde{Re}\frac{\partial \Sigma_T^{AA}(p^2)}{\partial p^2}|_{p^2=0}
         + \frac{\sin \theta_W}{\cos \theta_W} \frac{\tilde{Re}\Sigma_T^{ZA}(0)}{m_Z^2},
\end{equation}

\begin{equation}
\label{ct3}
\delta Z_{H^+H^+}= -\tilde{Re}\frac{\partial
\Sigma_T^{H^{\pm}}(p^2)}{\partial
p^2}|_{p^2=M_{H^{\pm}}^2},~~~\delta M_{H^{\pm}} = \tilde{Re}
\Sigma_T^{H^+}(M_{H^{\pm}}^2)-\delta b_{AA},
\end{equation}
where
\begin{equation}
\delta b_{AA}=  \frac{e}{m_W \sin \theta_W \sin 2 \beta}[\delta
T_{H}(\sin^3\beta\cos\alpha + \cos^3\beta\sin\alpha) + \delta
T_{h}(\cos^3\beta\cos\alpha - \sin^3\beta\sin\alpha)],
\end{equation}

\begin{equation}
\delta T_{H}= T_{H},~~~\delta T_{h}= T_{h}.
\end{equation}
The notation $\tilde{Re}$ appearing in Eqs.(\ref{ct1}),
(\ref{ct2}) and (\ref{ct3}), means taking the real part of the
loop integrals appearing in the self-energy.

\par
We take the fine structure constant at the $Z^0$-pole as input
parameter, Then we use the counter-term of the electric charge in
$\overline {DR}$ scheme expressed as\cite{count1, count2, eberl}
\begin{eqnarray}
\label{ecount}
   \delta Z_e &=&
   \frac{e^2}{6(4\pi)^2}
    \left\{ 4 \sum_f N_C^f e_f^2\left( \Delta+\log\frac{Q^2}{x_f^2} \right)+\sum_{\tilde{f}} \sum_{k=1}^2 N_C^f
    e_{f}^2 \left( \Delta+\log\frac{Q^2}{m^2_{\tilde{f}_k}} \right)     \right.       \nonumber  \\
  &&\left. + 4 \sum_{k=1}^2\left(\Delta+\log\frac{Q^2}{m^2_{\tilde{\chi}_k}}\right)
       +\sum_{k=1}^2\left( \Delta+\log\frac{Q^2}{m^2_{H_k^+}} \right)   \right.      \nonumber   \\
  &&\left.  - 22 \left(\Delta+\log\frac{Q^2}{m_W^2}\right)
  \right\},
\end{eqnarray}
where we take $x_f=m_Z$ when $m_f<m_Z$ and $x_t=m_t$. $e_f$ is the
electric charge of (s)fermion and
$\Delta=2/\epsilon-\gamma+\log4\pi$. $N_C^f$ is color factor,
which equal to 1 and 3 for (s)leptons and (s)quarks, respectively.

\par
The one-loop virtual corrections to $\gamma \gamma \to H^+H^-$ is
represented as
\begin{eqnarray}
\hat{\sigma}^{V}(\hat{s}, \gamma\gamma \to H^+H^-) = \frac{1}{16
\pi \hat{s}^2} \int_{\hat{t}_{min}}^{\hat{t}_{max}} d\hat{t}~ 2 Re
\overline{\sum} [(M^{V})^{\dagger} M^0],
\end{eqnarray}
where $\hat{t}_{max,min}= (M^2_{H^\pm}- \frac{1}{2} \hat{s}) \pm
\frac{1}{2}\sqrt{\hat{s}^2- 4 M^2_{H^\pm} \hat{s}} $, and the
summation with bar over head means the same operation as that
appeared in Eq.(\ref{folding}). $M^{V}$ is the renormalized
amplitude for virtual one-loop corrections. After renormalization
procedure, $\hat{\sigma}^{V}$ is UV-finite. Nevertheless, it still
contains the soft IR singularities. The IR singularity in the
$\hat{\sigma}^{V}$ is originated from the virtual photonic loop
correction, It can be cancelled by the contribution of the real
photon emission corrections. We shall discuss that in the
following subsection.

\par
\subsection{Real Photon Emission Corrections }

\par
\begin{figure}[htb]
\centering
\includegraphics{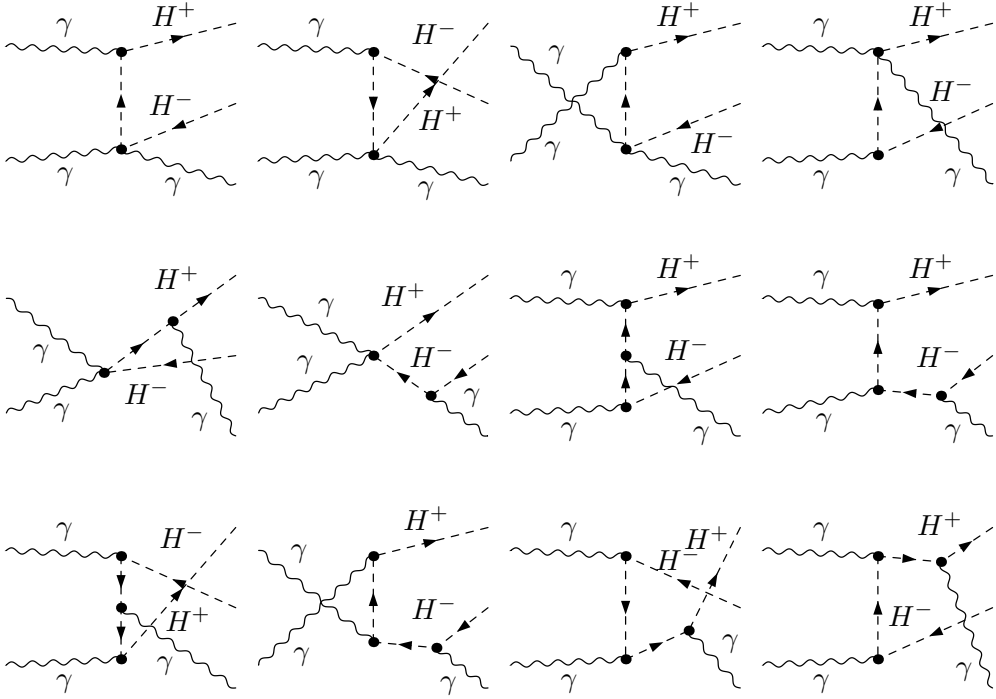}
\caption{The real photon emission diagrams for the subprocess
$\gamma \gamma \to H^+H^- \gamma $.} \label{fig:feyn_realphoton}
\end{figure}

We denote the real photon emission process as
\begin{equation}
\label{process2}
 \gamma (p_1) + \gamma(p_2) \to H^+(p_3) +
H^-(p_4) + \gamma (k) ,
\end{equation}
where $k = (k^0, \vec{k})$ is the four-momentum of the radiated
photon, and $p_1$, $p_2$, $p_3$ and $p_4$ are the four-momenta of
two initial photons and final charged Higgs pair $H^+H^-$,
respectively. The real photon emission Feynman diagrams for the
process $\gamma \gamma \to H^+H^- \gamma$ are displayed in
Fig.\ref{fig:feyn_realphoton}.
In our paper, we adopt the general phase-space-slicing
method\cite{PSS} to separate the soft photon emission singularity
from the real photon emission process. By using this method, the
bremsstrahlung phase space is divided into singular and
non-singular regions. Then the correction of the real photon
emission is broken down into corresponding soft and hard terms
\begin{equation}
\Delta \hat{\sigma}_{real} =\Delta \hat{\sigma}_{soft}+\Delta
\hat{\sigma}_{hard}
=\hat{\sigma}_0(\hat{\delta}_{soft}+\hat{\delta}_{hard}).
\end{equation}
In the c.m.s. frame, the radiated photon energy $k^0 =
\sqrt{|\vec{k}|^2 + m_{\gamma}^2}$ is called `soft' if $k^0 \leq
\Delta E_{\gamma}$ or `hard' if $k^0 > \Delta E_{\gamma}$. Here,
$m_{\gamma}$ is a small photon mass, which is used to regulate the
infrared divergences existing in the soft term. Although both
$\Delta \hat{\sigma}_{soft}$ and $\Delta \hat{\sigma}_{hard}$
depend on the soft photon cutoff $\Delta E_{\gamma}/E_b$, where
$E_b = \frac{\sqrt{\hat s}}{2}$ is the electron beam energy in the
c.m.s. frame, the real correction $\Delta \hat{\sigma}_{real}$ is
cutoff independent. In the calculation of soft term, we use the
soft photon approximation. Since the diagrams in
Fig.\ref{fig:feyn_realphoton} with real photon radiation from the
internal charge Higgs line or photon-charge Higgs vertex do not
lead to IR-singularity, we can neglect them in the calculation of
soft photon emission subprocesses (\ref{process2}) by using the
soft photon approximation method.  In this approach the
contribution of the soft photon emission subprocess is expressed
as\cite{COMS,Velt}
\begin{equation}
{\rm d} \Delta \hat{\sigma}_{{\rm soft}} = -{\rm d} \hat{\sigma}_0
\frac{\alpha_{{\rm ew}}e_{H^{+}}^2}{2\pi^2}
 \int_{|\vec{k}| \leq \Delta E_{\gamma}}\frac{{\rm d}^3 k}{2 k^0} \left[
 \frac{p_3}{p_3 \cdot k}-\frac{p_4}{p_4 \cdot k} \right]^2
\end{equation}
where the soft photon cutoff $\Delta E_{\gamma}$ satisfies $k^0
\leq \Delta E_{\gamma} \ll \sqrt{\hat{s}}$. The integral over the
soft photon phase space has been implemented in Ref.\cite{COMS},
then one can obtain the analytical result of the soft real photon
emission correction to $\gamma\gamma \to H^+H^-$.
\par
As mentioned above, the IR divergence of the virtual photonic
corrections can be exactly cancelled by that of soft real
correction. Therefore, $\Delta \hat{\sigma}_{{\rm vir} + {\rm
soft}}$, the sum of the  virtual and soft contributions, is
independent of the IR regulator $m_{\gamma}$. In the following
numerical calculations, we have checked the cancellation of IR
divergencies and verified that the total contributions of soft
photon emission and the virtual corrections are numerically
independent of $m_{\gamma}$. In addition, we present the numerical
verification of that the total one-loop level EW correction to the
cross section of $\gamma \gamma \to H^+H^-$, defined as $\Delta
\hat{\sigma}=\Delta \hat{\sigma}_{vir} +\Delta
\hat{\sigma}_{real}$, is independent of the cutoff $\Delta
E_{\gamma}$.
\par
Finally, we get an UV and IR finite correction $\Delta
\hat{\sigma}$:
$$
\Delta \hat{\sigma} = \Delta \hat{\sigma}_{vir} +\Delta
\hat{\sigma}_{real} =\hat{\sigma}_0 \hat{\delta}
$$
where
$\hat{\delta}=\hat{\delta}_{vir}+\hat{\delta}_{soft}+\hat{\delta}_{hard}$
is the one-loop relative correction.

\par
\subsection{Calculation of the Parent Process $e^+e^- \to \gamma\gamma \to H^+H^-$ }

\par
The total cross section of the parent process $e^+e^- \to
\gamma\gamma\to H^+H^-$ can be written as
\begin{eqnarray}
\label{integration}
\hat\sigma(s)= \int_{E_{0}/ \sqrt{s}} ^{x_{max}} d z \frac{d%
{\cal L}_{\gamma\gamma}}{d z} \hat{\sigma}(\gamma\gamma \to H^+H^-
\hskip 3mm
 at \hskip 3mm  \hat{s}=z^{2} s)
\end{eqnarray}
with $E_{0}=2m_{H^\pm}$, and $\sqrt{s}$($\sqrt{\hat{s}}$) being
the $e^{+}e^{-}$($\gamma\gamma$) center-of-mass energy.
$\frac{d\cal L_{\gamma\gamma}}{d z}$ is the distribution function
of photon luminosity, which is defined as:
\begin{eqnarray}
\frac{d{\cal L}_{\gamma\gamma}}{dz}=2z\int_{z^2/x_{max}}^{x_{max}}
 \frac{dx}{x} F_{\gamma/e}(x)F_{\gamma/e}(z^2/x)
\end{eqnarray}
For the initial unpolarized electrons and laser photon beams, the
energy spectrum of the back scattered photon is given
by\cite{photon spectrum}
\begin{eqnarray}
\label{structure}
F_{\gamma/e}=\frac{1}{D(\xi)}[1-x+\frac{1}{1-x}-
\frac{4x}{\xi(1-x)}+\frac{4x^{2}}{\xi^{2}(1-x)^2}]
\end{eqnarray}
where
\begin{eqnarray}
D(\xi)=\left(1-\frac{4}{\xi}-\frac{8}{{\xi}^2}\right)\ln{(1+\xi)}+\frac{1}{2}+
  \frac{8}{\xi}-\frac{1}{2{(1+\xi)}^2},
\end{eqnarray}
where $\xi=\frac{4E_0 \omega_0}{{m_e}^2}$, $m_{e}$ and $E_{0}$ are
the incident electron mass and energy, respectively, $\omega_0$ is
the laser-photon energy, and $x$ is the fraction of the energy of
the incident electron carried by the backscattered photon. In our
calculation, we choose $\omega_0$ such that it maximizes the
backscattered photon energy without spoiling the luminosity via
$e^{+}e^{-}$ pair creation. Then we have ${\xi}=2(1+\sqrt{2})$,
$x_{max}\simeq 0.83$, and $D(\xi)=1.8$.

\par
\section{Numerical results and discussion}

\par
We take the SM input parameters as $m_e=0.511~MeV$,
$m_\mu=105.66~MeV$, $m_\tau=1.777~GeV$, $m_Z = 91.188~GeV$, $m_W =
80.425~GeV$, $m_u = 66~MeV$, $m_c = 1.2~GeV$, $m_t = 178.1~GeV$,
$m_d = 66~MeV$, $m_s = 150~MeV$, $m_b = 4.7~ GeV$,
$\alpha_{ew}(m_Z^2)^{-1}|_{\overline{MS}}=127.918$\cite{pdg}.
There we use the effective values of the light quark masses ($m_u$
and $m_d$) which can reproduce the hadron contribution to the
shift in the fine structure constant
$\alpha_{ew}(m_Z^2)$\cite{leger}.

\par
The MSSM parameters are determined by using FormCalc package with
following input parameters\cite{FormCalc}:

\par
(1) The input parameters for the Higgs sector are the charged
Higgs mass $M_{H^\pm}$ and $\tan \beta$. The output masses of
other Higgs bosons are fixed by taking into account the
significant radiative corrections (Actually the results are almost
invariant quantitatively no matter which relation(tree or 2-loop
level) we use, since those masse values of Higgs bosons are only
adopted in the calculation of loop integrals.).

\par
(2) The input parameters for the chargino and neutralino sector
are the gaugino mass parameters $M_1$, $M_2$ and the Higgsino-mass
parameter $\mu$. We adopt the grand unification theory(GUT)
relation $M_1 = (5/3)\tan^2 \theta_W M_2$ for
simplification\cite{mssm-2}.

\par
(3) For the sfermion sector, we assume
$M_{\tilde{Q}}=M_{\tilde{U}}=M_{\tilde{D}}=M_{\tilde{E}}=M_{\tilde{L}}=M_{SUSY}$
and take the soft trilinear couplings for sfermions $\tilde{q}$
and $\tilde{l}$ being equal, i.e., $A_q=A_l=A_f$.

\par
Except above SM and MSSM input parameters, we have to input some
other parameters in our numerical calculations, for example, the
IR regularization parameter $m_{\gamma}$ and the soft cutoff
$\Delta E_{\gamma} /E_b$. In our following numerical calculations,
we take $\Delta E_{\gamma} /E_b=10^{-4}$ and
$m_{\gamma}=10^{-5}~GeV$, if there is no other statement. As we
know, the final results should be independent on the IR regulator
$m_{\gamma}$ and the soft cutoff $\Delta E_{\gamma} /E_b$. For
demonstration, we present the Fig.3, which shows the corrections
to the cross section of the subprocess $\gamma \gamma \to H^+H^-$
versus the soft cutoff $\Delta E_{g} /E_b$ in conditions of
$M_{H^{\pm}}= 250~GeV$, $\sqrt{\hat{s}}=1000~GeV$ and the input
parameters in $Set~1$(see below). The dashed, dotted and solid
lines correspond to $\Delta \hat{\sigma}_{vir+soft}$, $\Delta
\hat{\sigma}_{hard}$ and the total one-loop electroweak correction
$\Delta \hat{\sigma}$, respectively. As shown in this figure, the
full one-loop EW correction $\Delta \hat{\sigma}$ is independent
of the soft cutoff $\Delta E_\gamma/E_b$ as $\Delta E_\gamma /E_b$
running from $10^{-5}$ to $10^{-2}$, although both $\Delta
\hat{\sigma}_{vir+soft}$ and $\Delta \hat{\sigma}_{hard}$ depend
on the soft cutoff strongly.

\begin{figure}[htb]
\centering
\includegraphics[scale=0.5]{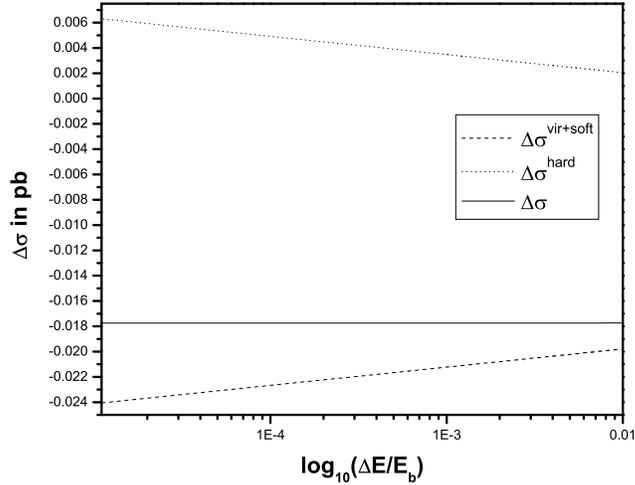}
\caption{The full one-loop corrections to the subprocess $\gamma
\gamma \to H^+H^-$ as the functions of the soft cutoff $\Delta
E/E_b$.}
\end{figure}

\par
In order to show the numerical results and discuss the effects of
the radiative corrections to the subprocess of $\gamma \gamma \to
H^+H^-$ quantitatively, we choose the following four typical input
data sets:

\begin{itemize}
\item[$Set~1$:] $\tan\beta=4$, $M_{H^{\pm}}=250 ~~or~~ 500$ GeV,
$M_{SUSY}=200$ GeV, $\mu=600$ GeV, $M_2=200$ GeV and $A_f=400$
GeV.
 \item[$Set~2$:] $\tan\beta=20$, $M_{H^{\pm}}=250 ~~or~~ 500$ GeV, $M_{SUSY}=400$
GeV, $\mu=1000$ GeV, $M_2=200$ GeV and $A_f=800$ GeV.
 \item[$Set~3$:]
$\tan\beta=40$, $M_{H^{\pm}}=250~~ or~~ 500$ GeV, $M_{SUSY}=200$
GeV, $\mu=200$ GeV, $M_2=1000$ GeV and $A_f=300$ GeV.
\end{itemize}

With the input parameters $\tan\beta$, $M_{H^{\pm}}$, $M_{SUSY}$,
$\mu$, $M_2$ and $A_f$ in one of the above data sets, we can
obtain all the masses of supersymmetric particles by using package
FormCalc\cite{FormCalc}. The input data $Set1$(or $Set2$) with
small(or mediate) $\tan \beta$, makes the gaugino-like case with
lighter(or heavier) sfermions, while the input data $Set3$ with
larger $\tan \beta$ induces higgsino-like case.

\par
We also give the results at the SPS1a' point from the SPA
project\cite{SPA}. The fundamental SUSY parameters in SPA project
are compatible with all available precision data and actual mass
and cosmological bounds. The SPA convention parameters are defined
in the $\overline {DR}$ scheme at the scale of $Q = 1~TeV$. A
translation from these parameters to our on-mass-shell definition
can be performed by subtracting the corresponding counter terms,
i.e. ${\cal P}^{\rm OMS}={\cal P}(Q)-\delta{\cal P}(Q)$. Then we
get the pole mass of charged Higgs as $M_{H^{\pm}}=438.6 GeV $.
For all other parameters that do not enter in the tree-level
calculations, either $\overline {DR}$ or OMS value can be used,
since their difference is of higher order.

\begin{figure}[htbp]
\centering
\includegraphics[scale=0.4]{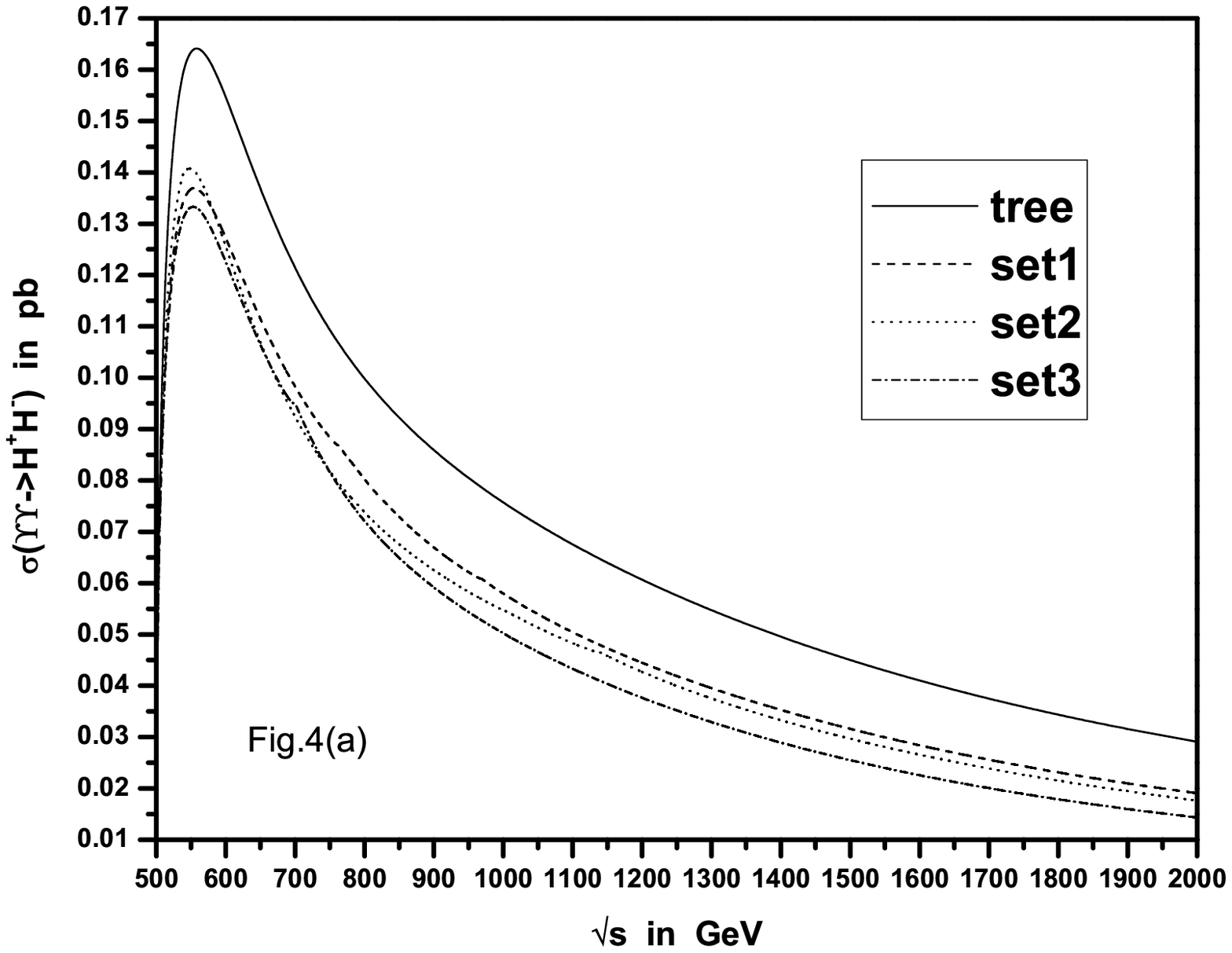}
\includegraphics[scale=0.4]{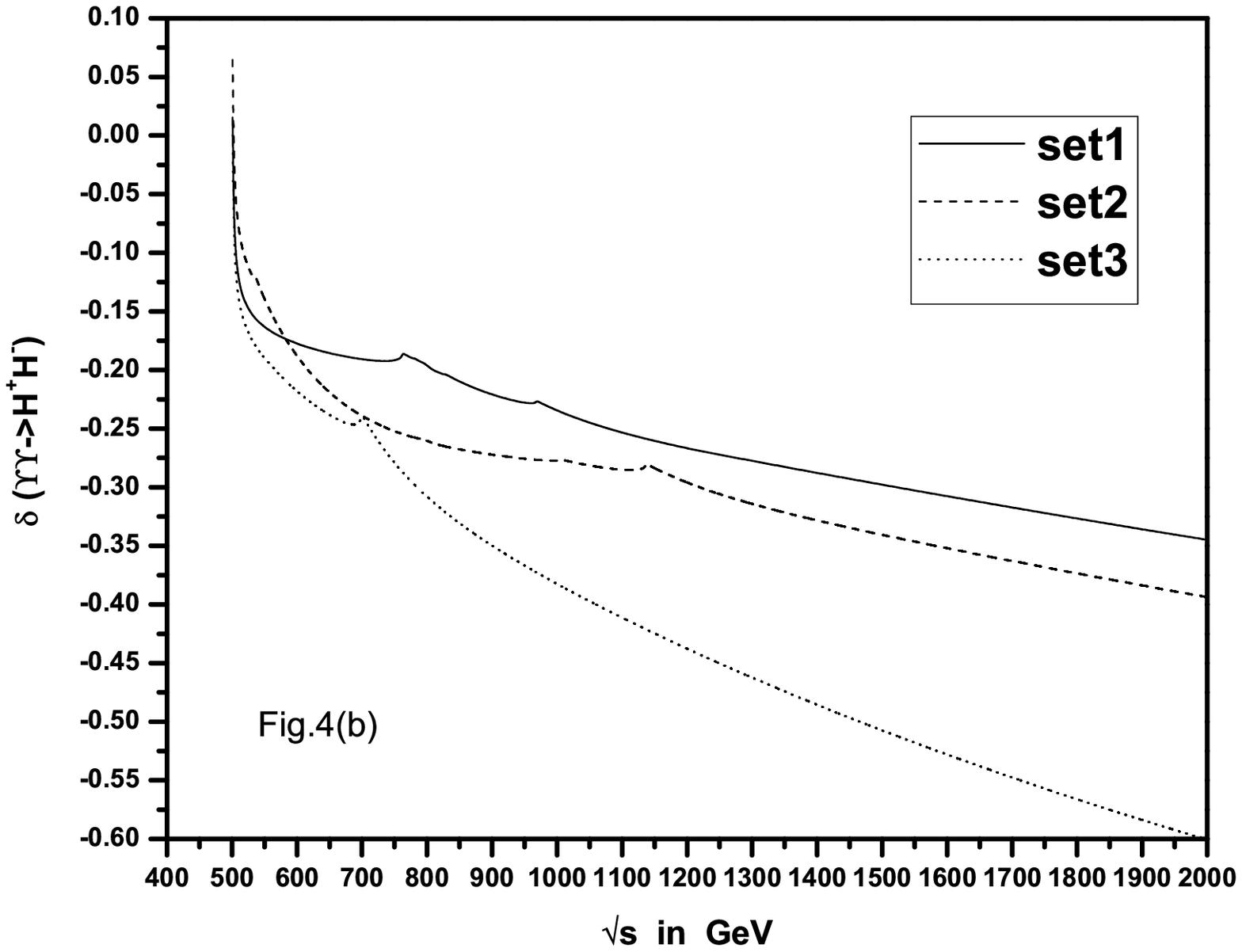}
\includegraphics[scale=0.4]{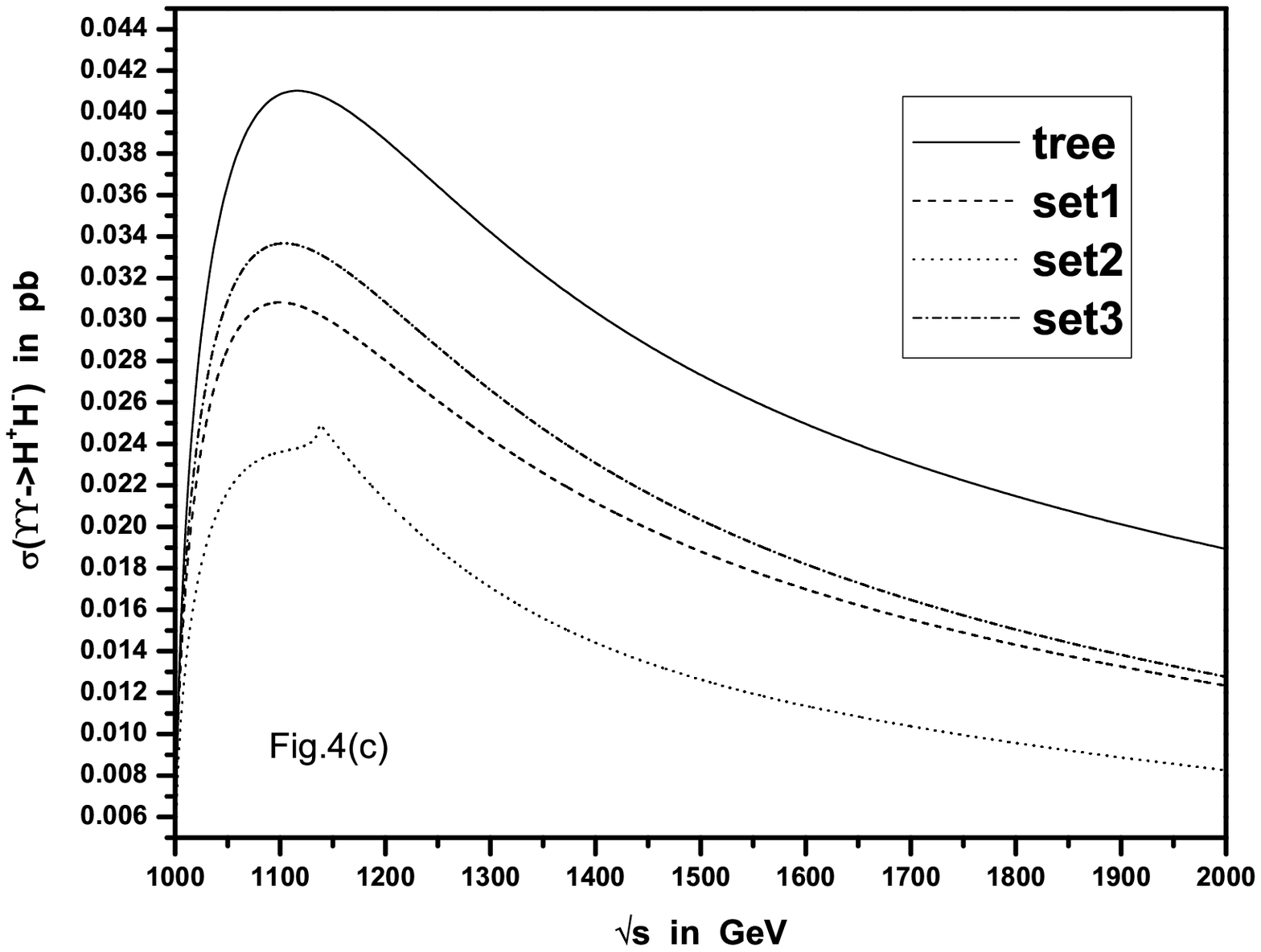}
\includegraphics[scale=0.4]{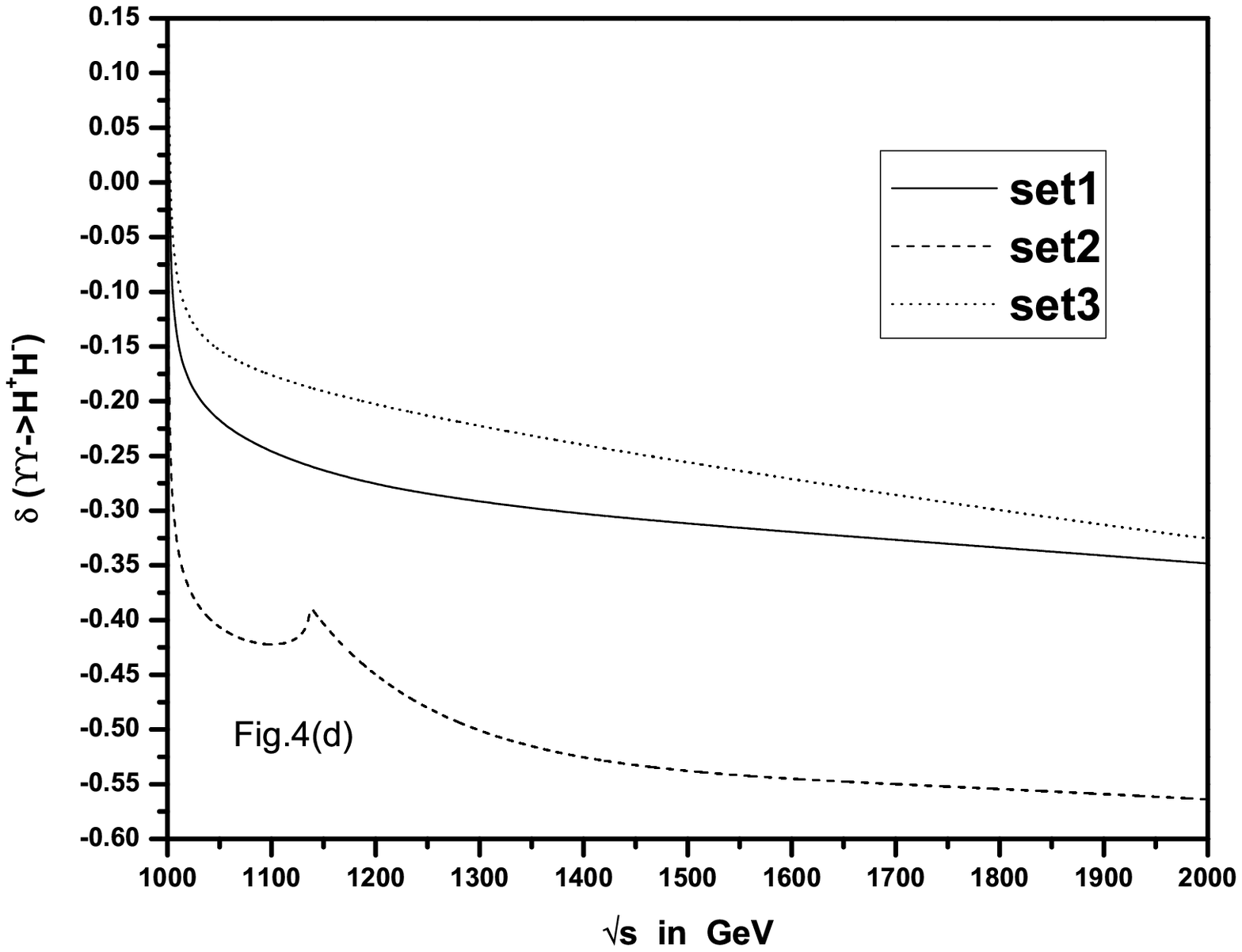}
\caption{The Born and the full one-loop level electroweak
corrected cross sections of the subprocess $\gamma \gamma \to
H^+H^-$ versus c.m.s. energy $\sqrt{\hat{s}}$ are plotted in
Fig.4(a)($M_{H^{\pm}} = 250 GeV$) and Fig.4(c)($M_{H^{\pm}} = 500
GeV$)). The corresponding relative corrections as the functions of
the c.m.s energy $\sqrt{\hat{s}}$ are shown in Fig.4(b) and
Fig.4(d), respectively.}
\end{figure}

\par
The Born and the full one-loop level electroweak corrected cross
sections for the subprocess $\gamma \gamma \to H^+H^-$ as the
functions of c.m.s. energy of $\gamma \gamma$ collision with above
three input data sets are displayed in Fig.4(a) with $M_{H^{\pm}}
= 250~GeV$ and in Fig.4(c) with $M_{H^{\pm}} = 500~GeV$,
respectively. The corresponding relative corrections are depicted
in Fig.4(b) and Fig.4(d). We can see that when $\sqrt{\hat s}\sim
558~(1118)$ GeV, the tree level cross section reaches the maximal
value 0.164 (0.041) pb in Fig.4(a) (Fig.4(c)). But its maximum
value is shifted to 0.141 (0.337) pb after including the one-loop
SUSY EW corrections. On the curve for the input data $Set~1$ in
Fig.4(b) there exist small resonance spikes in the regions around
the vicinities of $\sqrt{\hat{s}} \sim 2 m_{\tilde{t}_1}\sim
762.5~GeV$ and $\sqrt{\hat{s}}\sim 2m_{\tilde{t}_2}\sim
968.3~GeV$. On the curve for input data $Set~2$ the resonance peak
is located at $\sqrt{\hat{s}}\sim 2m_{\tilde{t}_2}\sim
1137.1~GeV$. For the curves of the input data $Set~3$, the
resonance effect can be seen around the position of
$\sqrt{\hat{s}}\sim 2m_{\tilde{\tau}_1}\sim 715.4~GeV$. On the
curves for the input data $Set~2$ in both Fig.4(c) and Fig.4(d),
we can see the resonance spikes at $\sqrt{\hat{s}}\sim
2m_{\tilde{t}_2}\sim 1137.1~GeV$. Both Fig.4(b) and Fig.4(d) show
that the relative corrections have their maximal values at the
position near the threshold energies and then decrease
quantitatively with the increment of $\sqrt{\hat s}$. At the
position of colliding energy $\sqrt{\hat s}=2~TeV$ shown in
Fig.4(b) (Fig.4(d)), the relative electroweak correction can reach
$-34.5\%~(-34.8\%)$, $-39.4\%~(-56.4\%)$ and $-60.1\%~(-32.6\%)$
for input data $Set~1$, $Set~2$ and $Set~3$, respectively. We can
see from Fig.4(b) that the absolute relative corrections for the
input data $Set~3$ can be rather large, and are generally larger
than the corresponding ones for the input data $Set~1$ and
$Set~2$. That is because in the input data $Set~3$ we have a very
small sbottom mass $m_{\tilde{b}_1}=76.67~GeV$. While Fig.4(d)
shows that the absolute relative corrections for the input data
$Set~2$ are the largest among all the three input data sets, since
there the conditions of $ m_{\tilde{l}_i,\tilde{U}_i,\tilde{D}_i }
\sim M_{H^{\pm}} = 500~GeV(i=1,2)$ are satisfied.

\begin{figure}[htbp]
\centering
\includegraphics[scale=0.5]{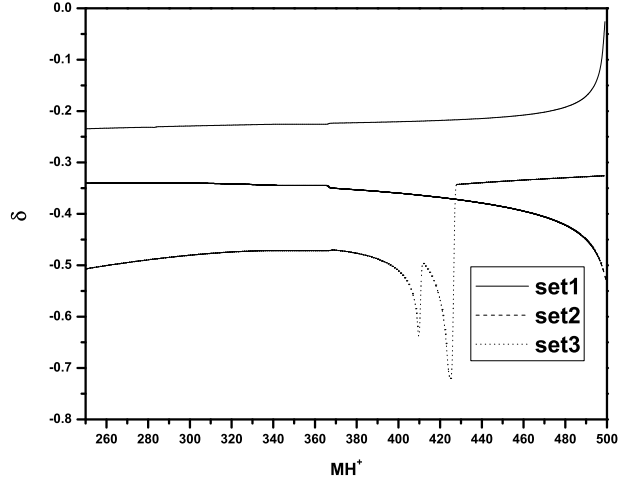}
\caption{The full one-loop relative electroweak corrections for
the subprocess $\gamma \gamma \to H^+H^-$ as the functions of the
charged Higgs mass $M_{H^{\pm}}$.}
\end{figure}

\par
In Fig.5 we present the full one-loop relative electroweak
corrections for the subprocess $\gamma \gamma \to H^+H^-$ as the
functions of the charged Higgs mass $M_{H^{\pm}}$ with input data
$Set~1$, $Set~2$ and $Set~3$ separately. The corresponding
collision energies in the c.m.s. of incoming $\gamma\gamma$,
$\sqrt {\hat s}$, are $1000~ GeV$, $1500~GeV$ and $2000~GeV$,
respectively. As shown in the figure all the curves are less
sensitive to $M_{H^{\pm}}$, except in the region of
$M_{H^{\pm}}>390~GeV$. For the curve with the input data $Set~3$,
the resonance effect can be seen around the positions of
$M_{H^{\pm}}\sim m_{\tilde{t}_1} + m_{\tilde{b}_2}\sim 410.6~GeV$
and $M_{H^{\pm}}\sim m_{\tilde{t}_2} + m_{\tilde{b}_1}\sim
426.2~GeV$. In this figure the curve for the input data $Set~1$
shows that the relative correction is almost stable except in the
energy region approaching the threshold $\sqrt{\hat{s}} \sim 2
M_{H^{\pm}}\sim 1000~GeV$. However, the curve for the input data
set $Set~3$ varies sharply in the region of
$380~GeV<M_{H^{\pm}}<440~GeV$ due to the resonance effects.

\begin{figure}[htbp]
\centering
\includegraphics[scale=0.5]{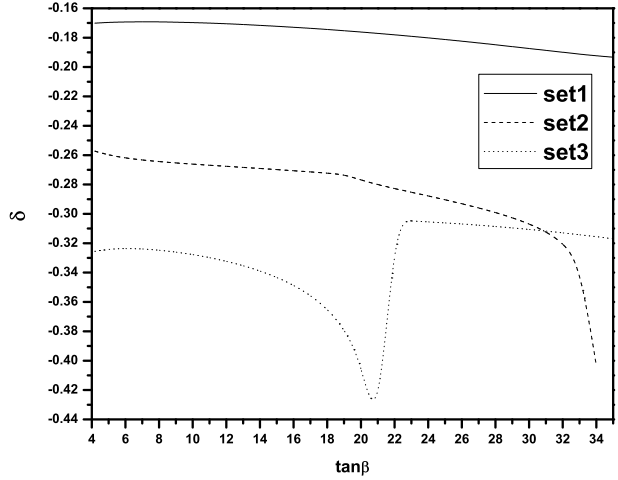}
\caption{The full one-loop relative electroweak corrections for
the subprocess $\gamma \gamma \to H^+H^-$ as the functions of
$\tan\beta$.}
\end{figure}

\par
In each figure of Fig.6, Fig.7 and Fig.8, we take the input
parameter sets as: $Set~1$, $Set~2$ and $Set~3$, except the charge
Higgs mass being $200~GeV$, $300~GeV$ and $500~GeV$, the colliding
energy being $500~GeV$, $1000~GeV$ and $2000~GeV$, respectively.
In Fig.6 we present the full one-loop relative electroweak
corrections for the subprocess $\gamma \gamma \to H^+H^-$ as the
functions of the ratio of the vacuum expectation values
$\tan\beta$. We can see from the curves for input data $Set~1$ and
$Set~2$ that the relative corrections decrease slowly with the
increment of $\tan\beta$ except the curve for $Set~2$ in the
region of $\tan\beta>32$. For the curve of the input data $Set~3$,
the resonance effect can be seen around the position of $\tan\beta
\sim 21$ where the condition of $M_{H^{\pm}}\sim m_{\tilde{t}_2} +
m_{\tilde{b}_1}\sim 500~GeV$ is satisfied.

\begin{figure}[htbp]
\centering
\includegraphics[scale=0.5]{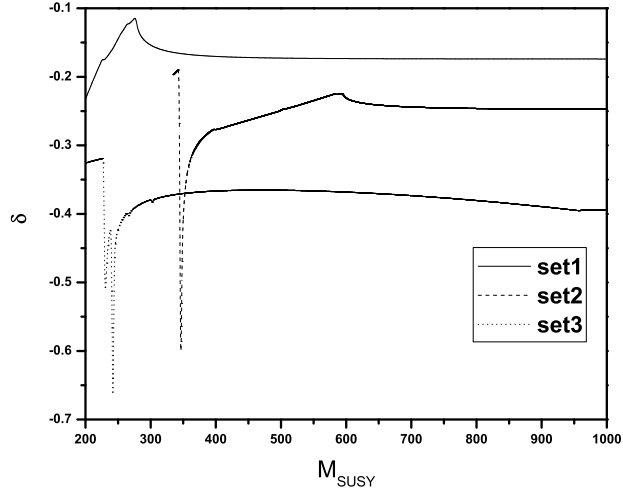}
\caption{The full one-loop relative electroweak corrections for
the subprocess $\gamma \gamma \to H^+H^-$ as the functions of
$M_{SUSY}$.}
\end{figure}

\par
In Fig.7 we present the full one-loop relative electroweak
corrections for the subprocess $\gamma \gamma \to H^+H^-$ as the
functions of $M_{SUSY}$. We can see that the relative corrections
are not sensitive to $M_{SUSY}$ except at the vicinities of the
resonance points. For the curve of the input data $Set~1$, the
resonance effect can be seen around the position of $M_{SUSY}\sim
276~GeV$ where $\sqrt{\hat{s}} \sim 2 m_{\tilde{t}_1}\sim
500~GeV$. For the curve of the input data $Set~2$, the resonance
effect can be seen around the positions of $M_{SUSY}\sim 346$ GeV
(where $M_{H^{\pm}}\sim m_{\tilde{t}_1} + m_{\tilde{b}_1}\sim
300~GeV$) and $M_{SUSY}\sim 594~GeV$ (where $\sqrt{\hat{s}} \sim 2
m_{\tilde{t}_1}\sim 1000~GeV$). For the curve of the input data
$Set~3$, the resonance effect can be seen around the positions of
$M_{SUSY}\sim 230~GeV$ (where $M_{H^{\pm}}\sim m_{\tilde{t}_2} +
m_{\tilde{b}_1}\sim 500~GeV$) and $M_{SUSY}\sim 242~GeV$ (where
$M_{H^{\pm}}\sim m_{\tilde{t}_1} + m_{\tilde{b}_2}\sim 500~GeV$).

\begin{figure}[htbp]
\centering
\includegraphics[scale=0.5]{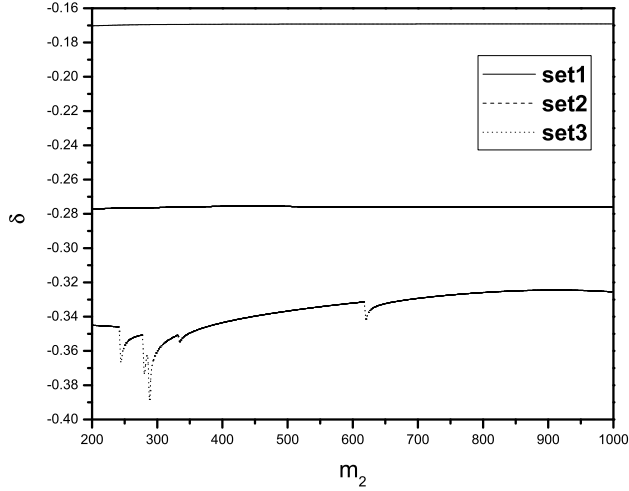}
\caption{The full one-loop relative electroweak corrections for
the subprocess $\gamma \gamma \to H^+H^-$ as the functions of
$M_2$.}
\end{figure}

\par
In Fig.8 we present the full one-loop relative electroweak
corrections to the subprocess $\gamma \gamma \to H^+H^-$ as the
functions of $M_2$. We can see that the relative corrections are
not sensitive to $M_2$ except the curve for the input data
$Set~3$. On the curve of the input data $Set~3$, there are five
resonance points which come from the fact that the masses of some
chargino and neutralino are lighter than $M_{H^{\pm}}$. The
resonance effect can be seen around the positions of $M_2\sim 244$
GeV (where $M_{H^{\pm}}\sim m_{\tilde{\chi}^+_2} +
m_{\tilde{\chi}^0_3}\sim 500~GeV$), $M_2\sim 280$ GeV (where
$M_{H^{\pm}}\sim m_{\tilde{\chi}^+_2} + m_{\tilde{\chi}^0_2}\sim
500~GeV$), $M_2\sim 288$ GeV (where $M_{H^{\pm}}\sim
m_{\tilde{\chi}^+_1} + m_{\tilde{\chi}^0_4}\sim 500~GeV$),
$M_2\sim 334$ GeV (where $M_{H^{\pm}}\sim m_{\tilde{\chi}^+_2} +
m_{\tilde{\chi}^0_1}\sim 500~GeV$) and $M_2\sim 620$ GeV (where
$M_{H^{\pm}}\sim m_{\tilde{\chi}^+_1} + m_{\tilde{\chi}^0_3}\sim
500~GeV$). For the input data set $Set~1$ and $Set~2$, the
relative corrections are almost stable, the relative corrections
are about $-16.9\%$ and $-27.6\%$, respectively.

\begin{figure}[htbp]
\centering
\includegraphics[scale=0.4]{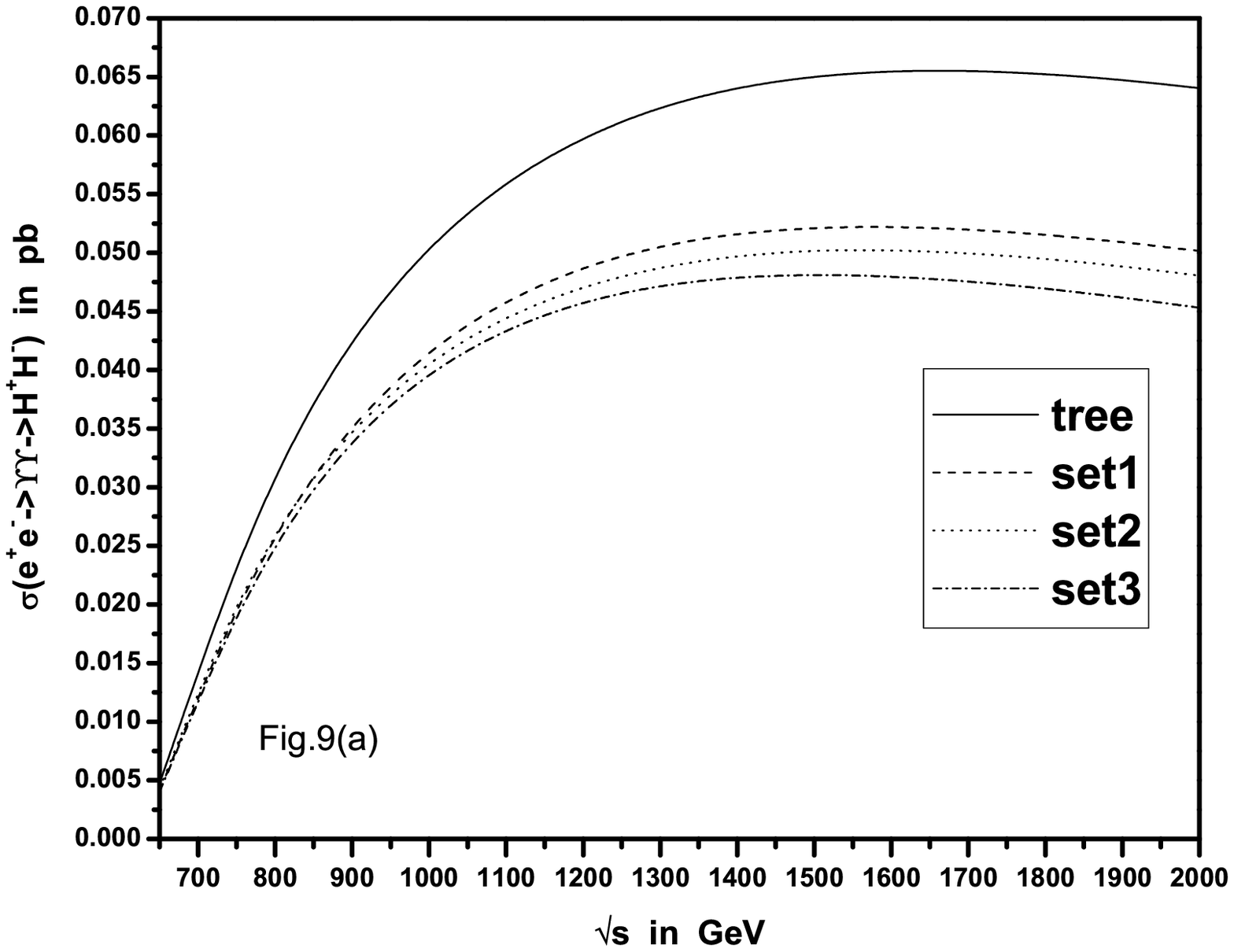}
\includegraphics[scale=0.4]{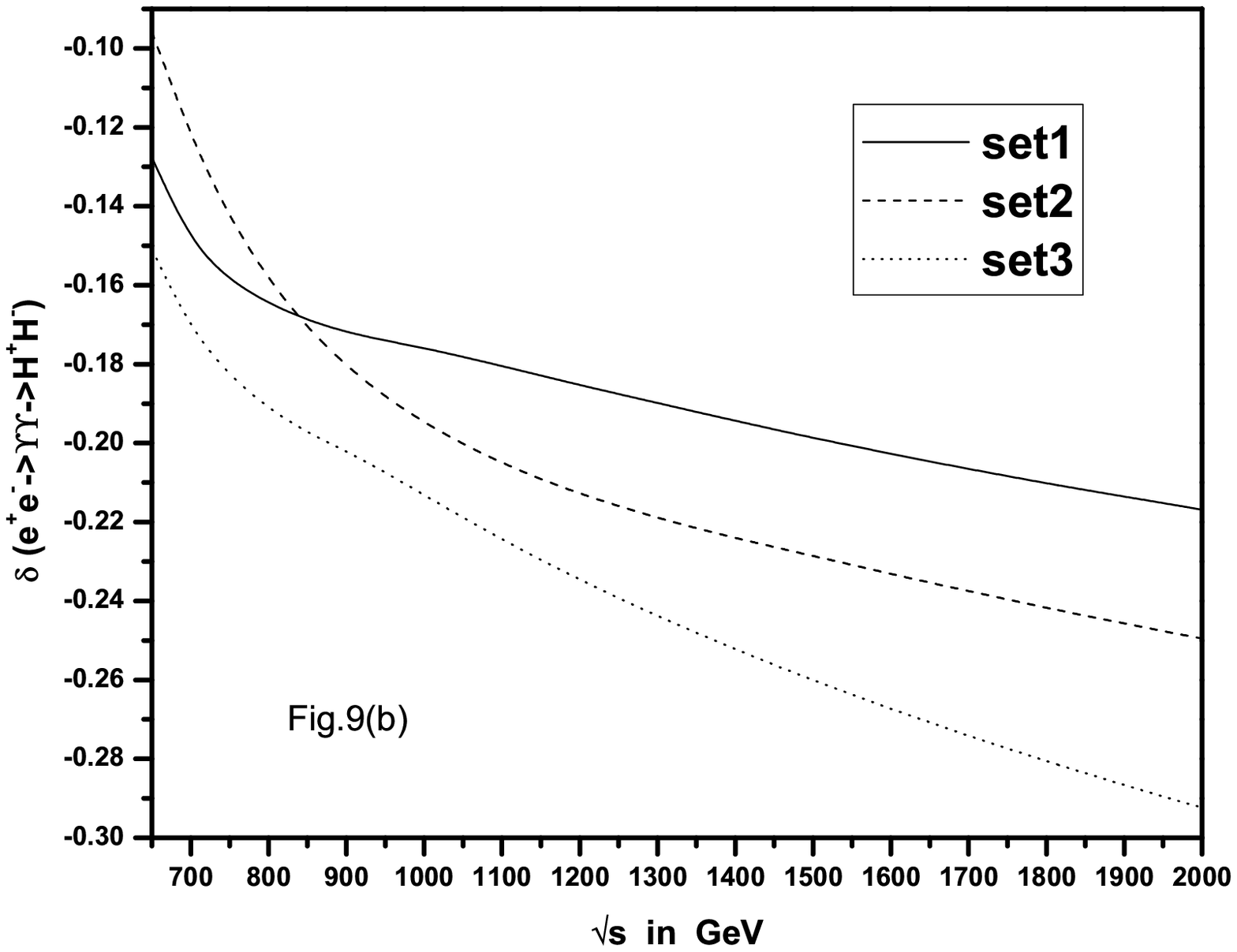}
\includegraphics[scale=0.4]{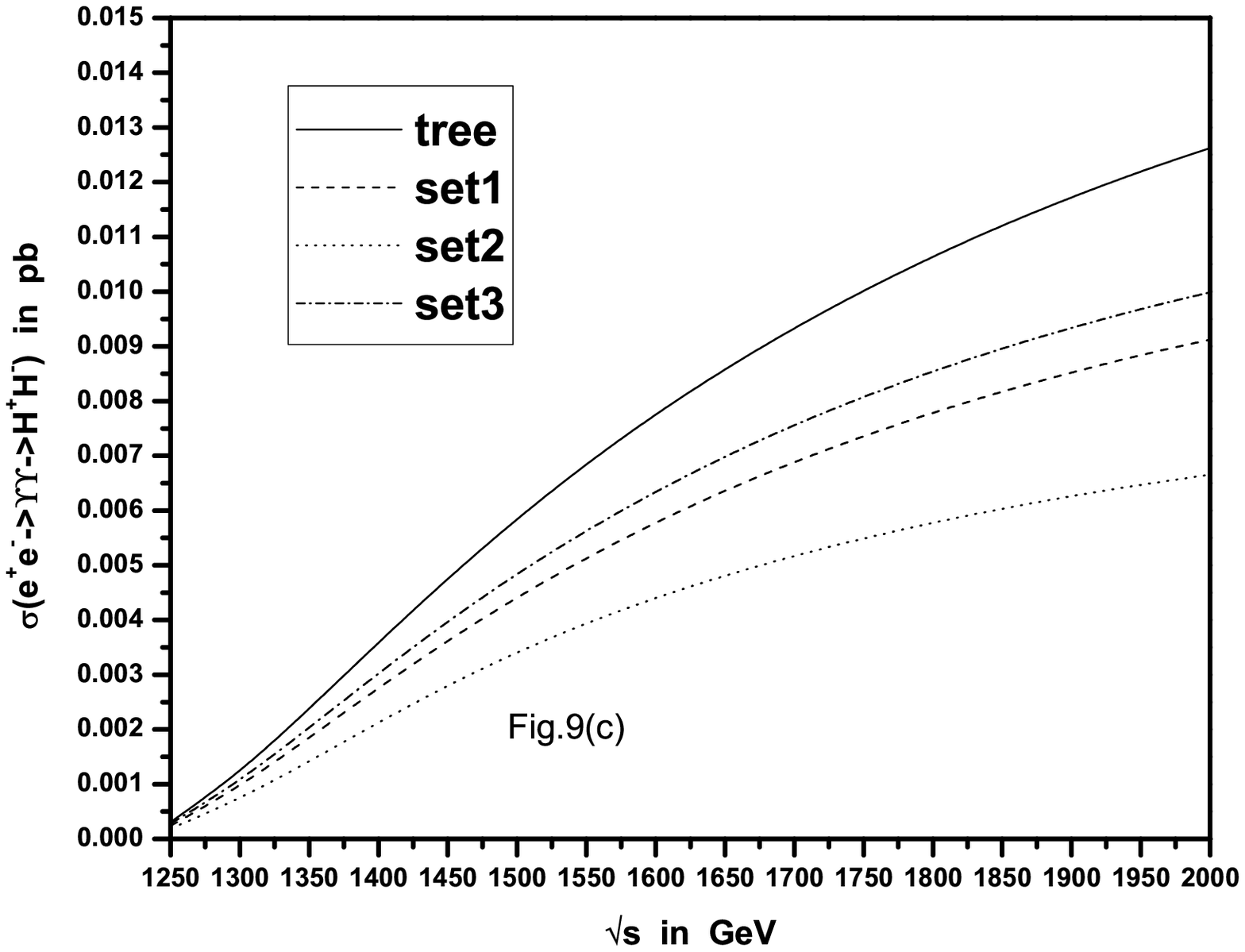}
\includegraphics[scale=0.4]{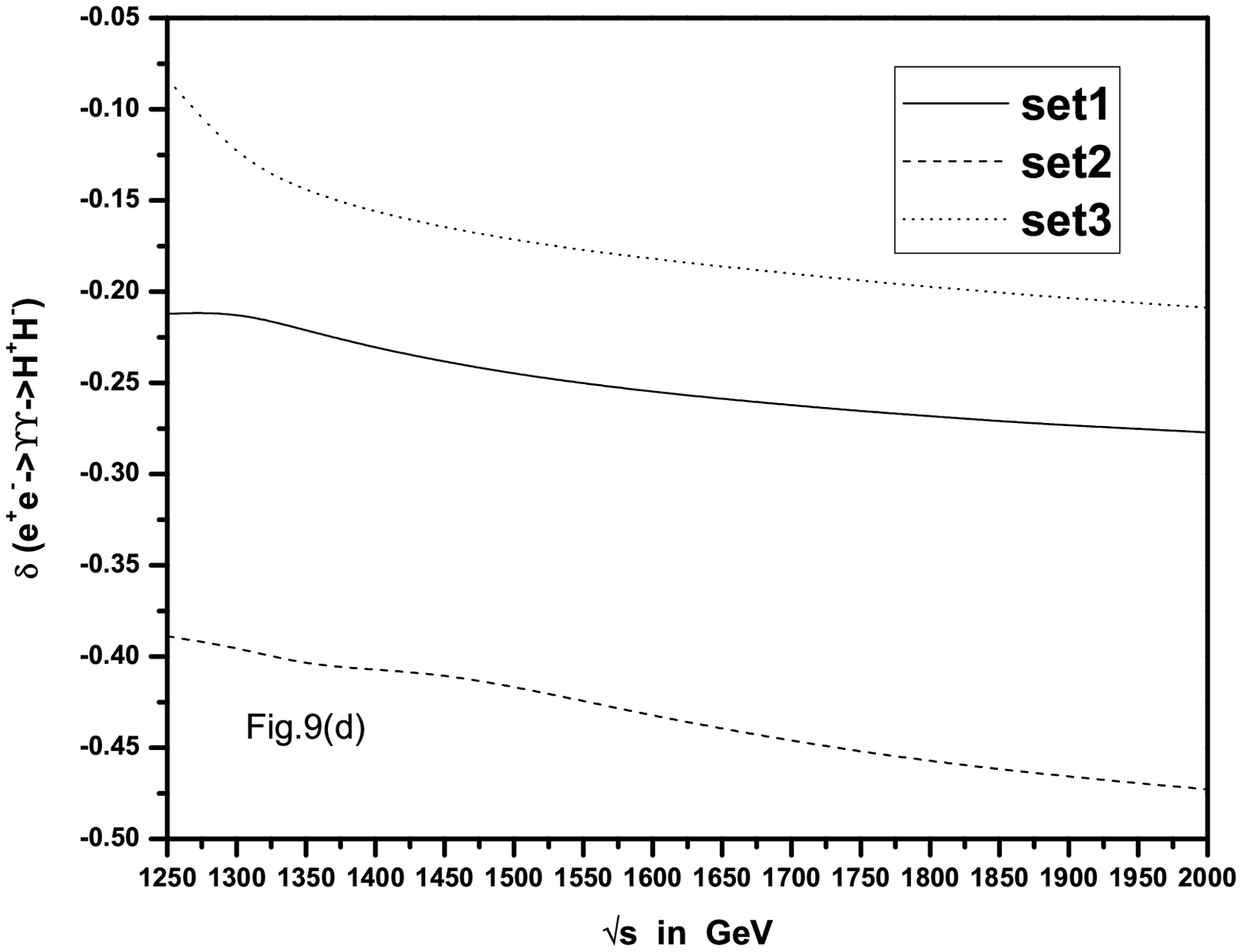}
\caption{The Born and the full one-loop level electroweak
corrected cross sections for the parent process $e^+e^-\to\gamma
\gamma \to H^+H^-$, are shown in Fig.9(a)($M_{H^{\pm}} = 250 GeV$)
and Fig.9(c)($M_{H^{\pm}} = 500 GeV$), respectively. The
corresponding relative corrections as the functions of the c.m.s
energy of the incoming electron-positron pair, are shown in
Fig.9(b) and Fig.9(d), separately.}
\end{figure}

\par
Fig.9(a) is the plot of the Born and the full one-loop level
electroweak corrected cross sections for the parent process
$e^+e^-\to\gamma \gamma \to H^+H^-$ versus the electron-positron
colliding energy with $M_{H^{\pm}} = 250 GeV$. At the position of
$\sqrt{s} \sim 1.5~TeV$ the cross sections reach their maximal
values, e.g., for the curve of the input data $Set~1$, the cross
section has the maximal value $52.2~fb$. When $\sqrt{s}>1.5~TeV$,
the cross sections decrease slowly with the increment of the
colliding $e^+e^-$ c.m.s energy. Fig.9(b) shows the corresponding
relative corrections as the functions of colliding $e^+e^-$
energy. We can see that the absolute relative corrections increase
obviously with the increment of $\sqrt{s}$. When $\sqrt {\hat s} =
2~TeV$, the relative correction can reach its maximal value
$-29.2\%$ for the curve of the input data $Set~3$. Fig.9(c) is the
plot of the Born and the full one-loop level electroweak corrected
cross sections for the parent process $e^+e^-\to\gamma \gamma \to
H^+H^-$ versus the electron-positron colliding energy with
$M_{H^{\pm}} = 500 GeV$. The cross sections increase with the
increment of the colliding c.m.s energy. When $\sqrt{s} = 2~TeV$,
the corrected cross section reaches its maximal value $10.9~fb$
for the input data $Set~3$. Fig.9(d) shows the corresponding
relative corrections as the functions of the colliding $e^+e^-$
energy. We can see that the absolute relative corrections increase
apparently with the increment of $\sqrt{s}$. When $\sqrt{s} =
2~TeV$, the relative correction can reach $-47.3\%$ for the input
data $Set~3$.

\begin{figure}[htbp]
\centering
\includegraphics[scale=0.4]{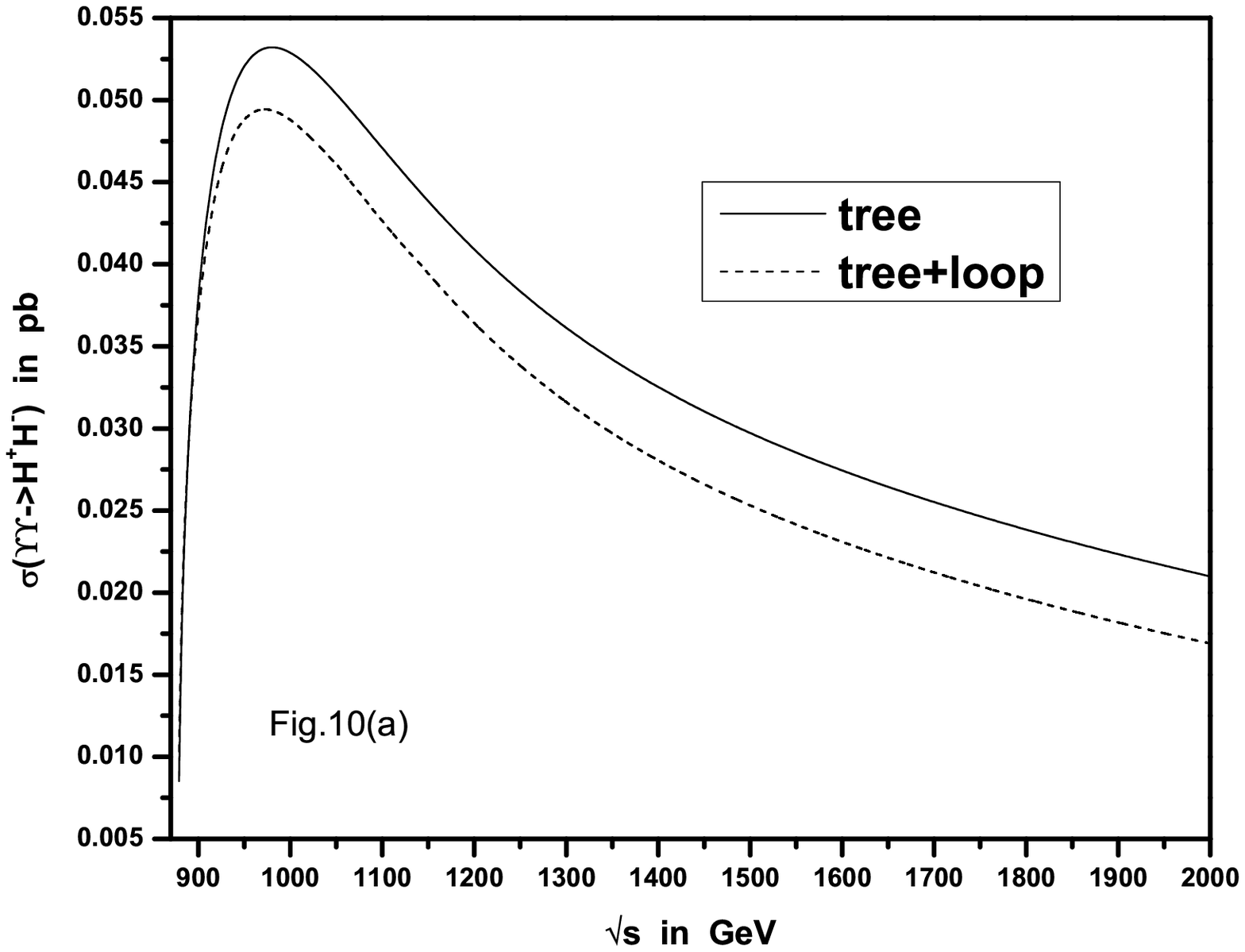}
\includegraphics[scale=0.4]{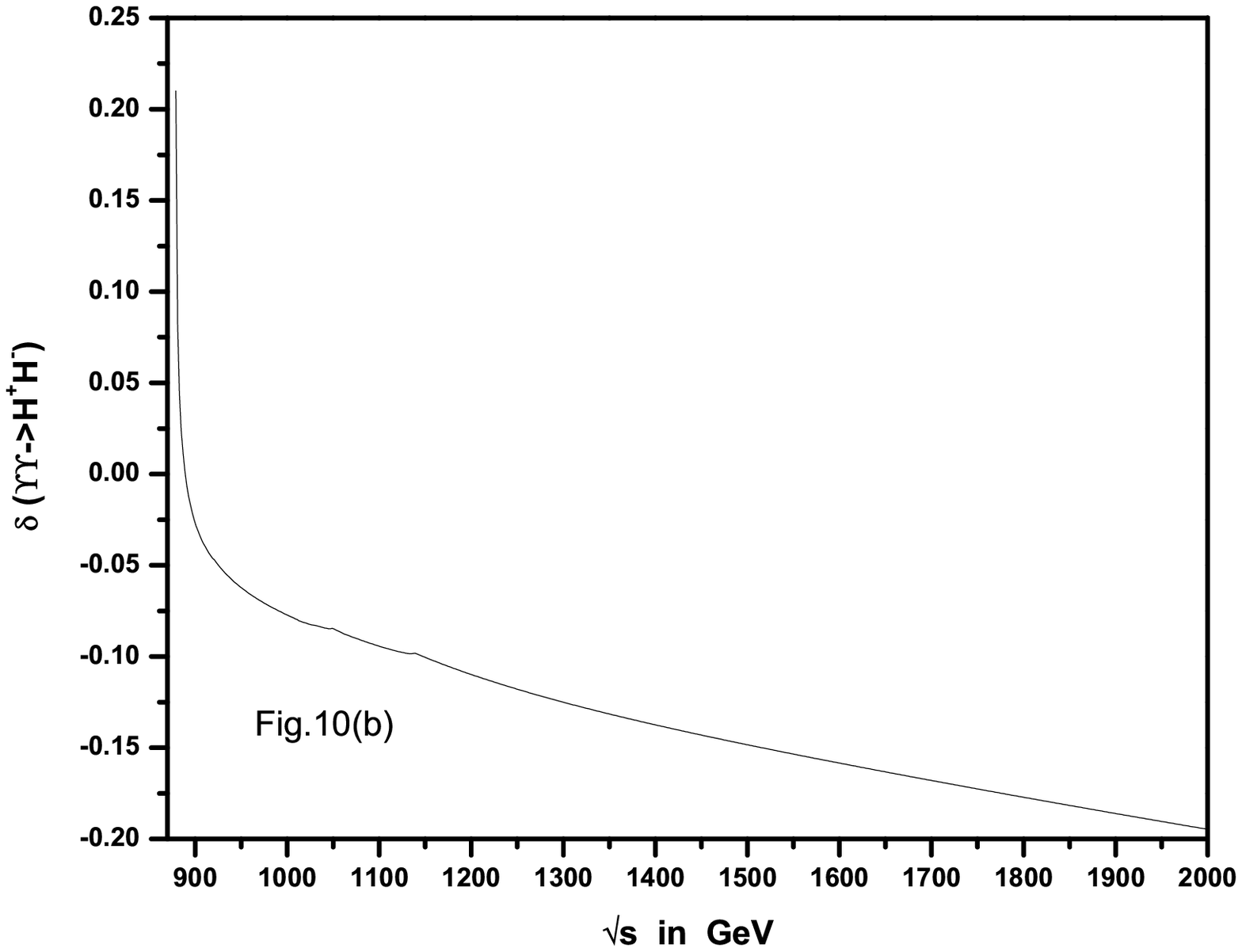}
\includegraphics[scale=0.4]{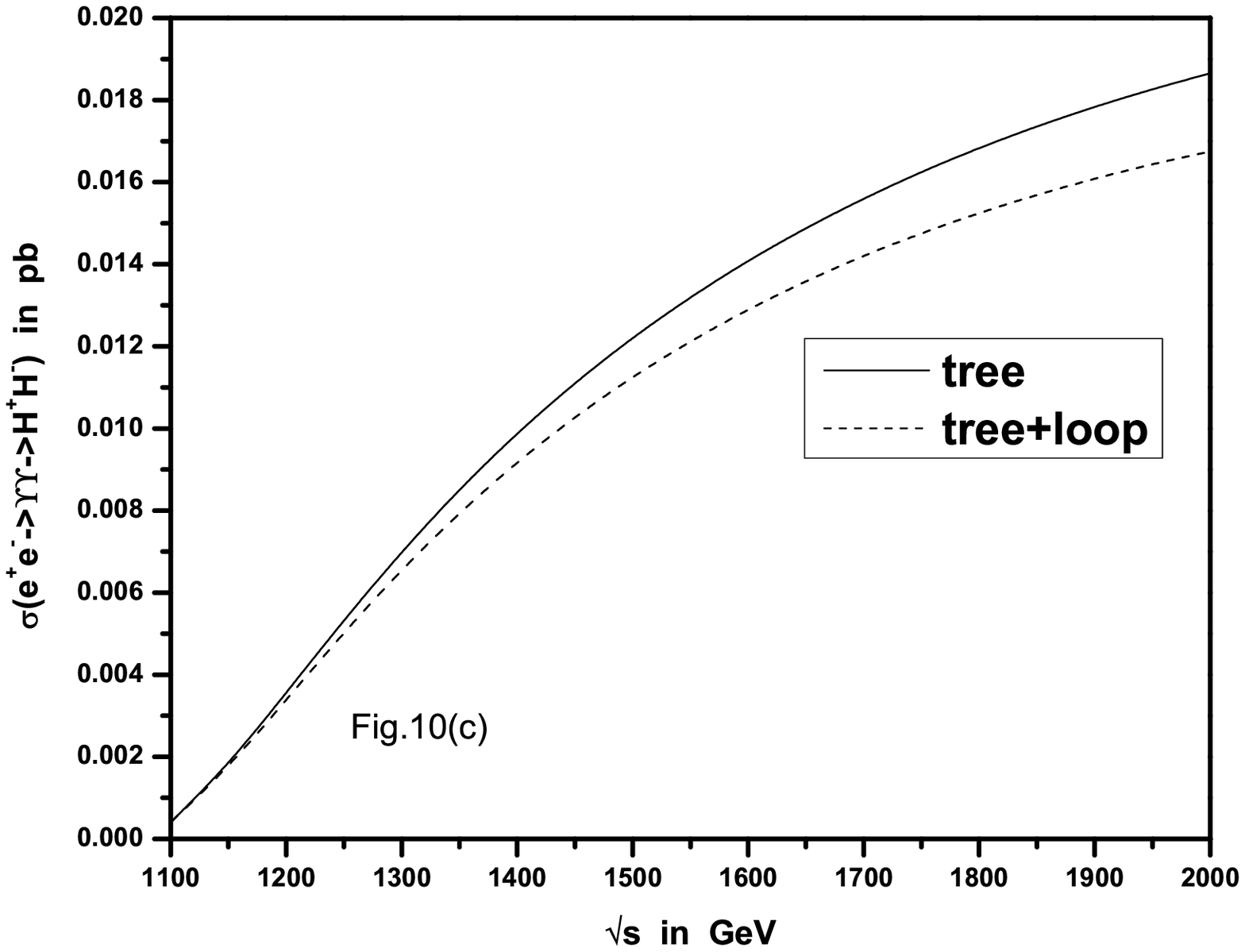}
\includegraphics[scale=0.4]{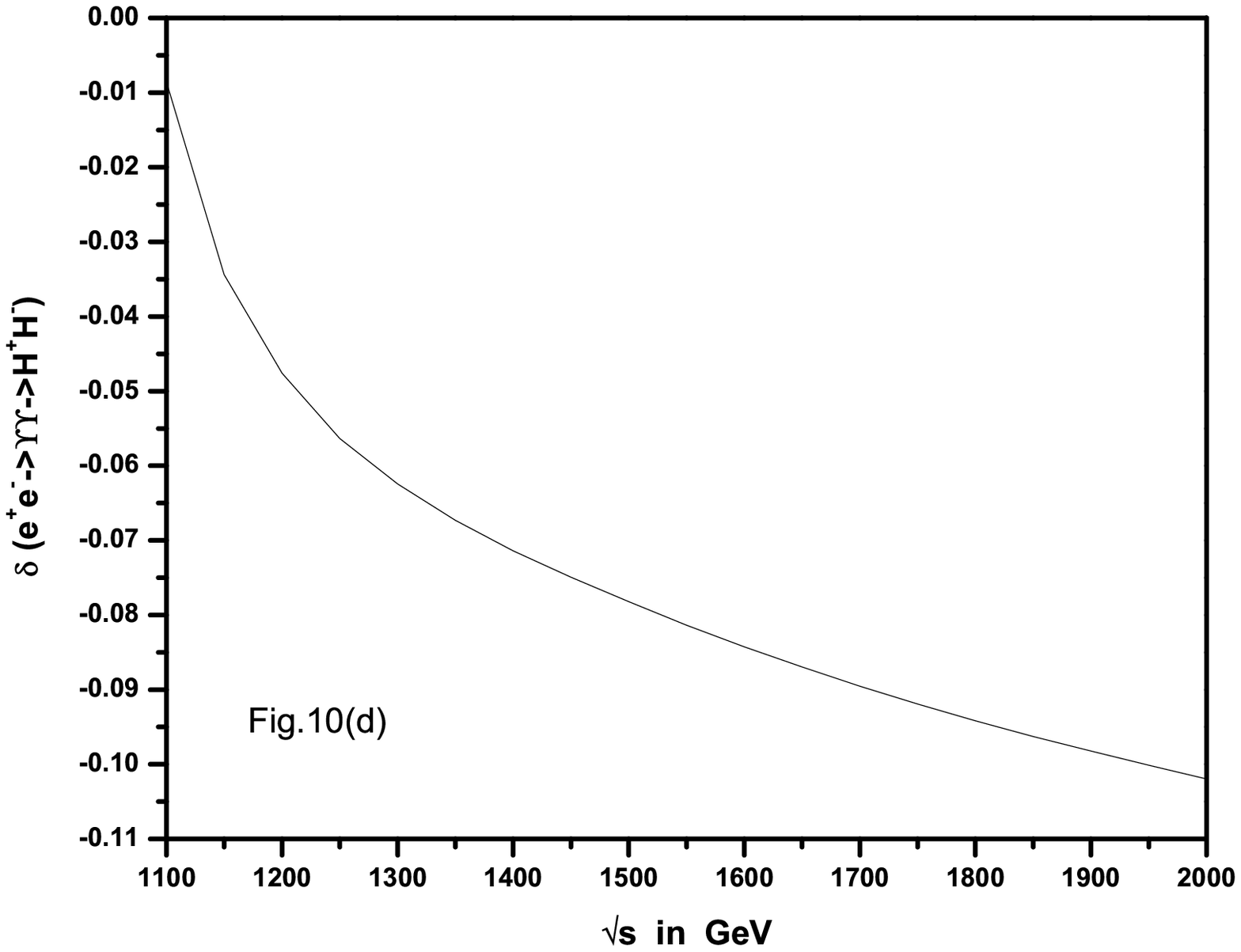}
\caption{The Born and the full one-loop level electroweak
corrected cross sections for the subprocess $\gamma \gamma \to
H^+H^-$(parent process $e^+e^- \to \gamma \gamma \to H^+H^-$), and
their corresponding relative corrections as the functions of the
c.m.s colliding energy $\sqrt{\hat{s}}(\sqrt{s})$ at the SPS1a'
point are shown in Fig.10(a) and Fig.10(b)(Fig.10(c) and
Fig.10(d)), respectively. }
\end{figure}

\par
Fig.10(a) and Fig.10(b) shows the Born and the full one-loop level
electroweak corrected cross sections and the corresponding
relative corrections for the subprocess $\gamma \gamma \to H^+H^-$
as the functions of the c.m.s energy $\sqrt{\hat{s}}$ at the
SPS1a' point. From Fig.10(a) we can see that the cross sections
decrease with the increment of the colliding c.m.s energy
$\sqrt{\hat{s}}$ when $\sqrt{\hat{s}} > 1~TeV$. At the point of
$\sqrt{\hat{s}} = 980~GeV$, the tree level and one-loop level
corrected cross sections are $53.2~fb$ and $49.4~fb$,
respectively. When $\sqrt{\hat{s}} = 2~TeV$, the tree level and
one-loop level corrected cross sections go down to $21.0~fb$ and
$15.7~fb$, respectively. In Fig.10(b) we can see that there are
two small peaks at the vicinities of $\sqrt{\hat{s}} \sim 2
m_{\tilde{c}_2}\sim 1048.3~GeV$ and $\sqrt{\hat{s}} \sim 2
m_{\tilde{t}_2}\sim 1137.4~GeV$. The absolute relative correction
increases with the increment of $\sqrt{\hat{s}}$. When
$\sqrt{\hat{s}}$ goes from $1~TeV$ to $2~TeV$, the relative
correction varies from $-7.76\%$ to $-19.5\%$. Fig.10(c) and
Fig.10(d) are the plots for the cross sections and the
corresponding relative corrections of the parent process
$e^+e^-\to\gamma \gamma \to H^+H^-$ as the functions of the c.m.s
energy of the incoming electron-positron pair, separately. From
Fig.10(c) we can see that the cross sections increase with the
increment of the colliding $e^+e^-$ c.m.s energy $\sqrt{s}$. When
$\sqrt{s} = 2~TeV$, the tree level and one-loop level corrected
cross sections are $18.7~fb$ and $16.8~fb$, respectively. In
Fig.10(d) we can see that the absolute relative correction
increases with the increment of $\sqrt{s}$. When $\sqrt{s}$ goes
from $1.1~TeV$ to $2~TeV$, the relative correction varies from
$-0.86\%$ to $-10.2\%$.

\par
\section{ Summary}
\par
In this paper, we present the calculation of the full one-loop
electroweak corrections to the subprocess $\gamma \gamma \to
H^+H^-$ and parent process $e^+e^-\to\gamma \gamma \to H^+H^-$ at
a linear collider in the MSSM. We analyze the dependence of the
relative corrections for the subprocess on colliding energy,
charged Higgs boson mass and several supersymmetric parameters. We
find that these corrections generally reduce the Born cross
sections and the relative corrections are typically few dozen
percent for both the subprocess and the parent process . With the
input data $Set~3$, the relative corrections to the subprocess are
obviously sensitive to $M_{H^{\pm}} $, $\tan\beta$, $M_{SUSY}$ and
$M_2$ in some parameter space due to the resonance effects.
However, with the input data $Set~1$ and $Set~2$, the relative
corrections to the subprocess are less sensitive to these
parameters comparing with the curves with input data $Set~3$. We
also give the numerical results at the SPS1a' point, it shows that
with $\sqrt{\hat{s}}$ varying from $1~TeV$ to $2~TeV$, the
relative correction to the subprocess runs from $-7.76\%$ to
$-19.5\%$. We conclude that the complete one-loop electroweak
corrections to both subprocess $\gamma \gamma \to H^+H^-$ and the
parent process $e^+e^-\to\gamma \gamma \to H^+H^-$ are generally
significant and should be considered in the precise analysis.

\par
\noindent{\large\bf Acknowledgments:} This work was supported in
part by the National Natural Science Foundation of China and and a
special fund sponsored by China Academy of Science.

\vskip 10mm

\begin{flushleft} {\bf Figure Captions} \end{flushleft}
\par
{\bf Figure 1} The leading order diagrams for the $\gamma\gamma
\to H^+H^-$ subprocess.

\par
{\bf Figure 2} The real photon emission diagrams for the
subprocess $\gamma \gamma \to H^+H^- \gamma $.

\par
{\bf Figure 3} The full one-loop corrections to the subprocess
$\gamma \gamma \to H^+H^- $ as the functions of the soft cutoff
$\Delta E/E_b$.

\par
{\bf Figure 4} The Born and the full one-loop level electroweak
corrected cross sections of the subprocess $\gamma \gamma \to
H^+H^-$ versus c.m.s. energy $\sqrt{\hat{s}}$ are plotted in
Fig.4(a)($M_{H^{\pm}} = 250 GeV$) and Fig.4(c)($M_{H^{\pm}} = 500
GeV$)). The corresponding relative corrections as the functions of
the c.m.s energy $\sqrt{\hat{s}}$ are shown in Fig.4(b) and
Fig.4(d), respectively.

\par
{\bf Figure 5} The full one-loop relative electroweak corrections
for the subprocess $\gamma \gamma \to H^+H^-$ as the functions of
the charged Higgs mass $M_{H^{\pm}}$.

\par
{\bf Figure 6} The full one-loop relative electroweak corrections
for the subprocess $\gamma \gamma \to H^+H^-$ as the functions of
$\tan\beta$.

\par
{\bf Figure 7} The full one-loop relative electroweak corrections
for the subprocess $\gamma \gamma \to H^+H^-$ as the functions of
$M_{SUSY}$.

\par
{\bf Figure 8} The full one-loop relative electroweak corrections
for the subprocess $\gamma \gamma \to H^+H^-$ as the functions of
$M_2$.

\par
{\bf Figure 9} The Born and the full one-loop level electroweak
corrected cross sections for the parent process $e^+e^-\to\gamma
\gamma \to H^+H^-$, are shown in Fig.9(a)($M_{H^{\pm}} = 250 GeV$)
and Fig.9(c)($M_{H^{\pm}} = 500 GeV$), respectively. The
corresponding relative corrections as the functions of the c.m.s
energy of the incoming electron-positron pair, are shown in
Fig.9(b) and Fig.9(d), separately.

\par
{\bf Figure 10} The Born and the full one-loop level electroweak
corrected cross sections for the subprocess $\gamma \gamma \to
H^+H^-$(parent process $e^+e^- \to \gamma \gamma \to H^+H^-$), and
their corresponding relative corrections as the functions of the
c.m.s colliding energy $\sqrt{\hat{s}}(\sqrt{s})$ at the SPS1a'
point are shown in Fig.10(a) and Fig.10(b)(Fig.10(c) and
Fig.10(d)), respectively.

\end{document}